\documentclass[12pt]{article}
\usepackage[dvipdfmx]{graphicx}
\usepackage{color,amssymb,amsmath,ascmac}
\usepackage{slashed}

\begin{document}

\newcommand{\lrf}[2]{ \left(\frac{#1}{#2}\right)}
\newcommand{\lrfp}[3]{ \left(\frac{#1}{#2} \right)^{#3}}
\newcommand{\vev}[1]{\left\langle #1\right\rangle}

\newcommand{\TeV}{\text{TeV}}
\newcommand{\GeV}{\text{GeV}}
\newcommand{\MeV}{\text{MeV}}
\newcommand{\keV}{\text{keV}}
\newcommand{\eV}{\text{eV}}

\begin{titlepage}
\begin{flushright}
UT-16-07\\
IPMU-16-0015
\end{flushright}
\vskip 3cm
\begin{center}
{\Large \bf
Diphoton Excess and Running Couplings}
\vskip 1.5cm
{
Kyu Jung Bae$^{(a)}$,
Motoi Endo$^{(a,b)}$, 
Koichi Hamaguchi$^{(a,b)}$, 
Takeo Moroi$^{(a,b)}$
}
\vskip 0.9cm
{\it $^{(a)}$ Department of Physics, University of Tokyo, Bunkyo-ku, Tokyo 113--0033, Japan \vspace{0.2cm}
\par
$^{(b)}$ Kavli Institute for the Physics and Mathematics of the Universe (Kavli IPMU), \\
University of Tokyo, Kashiwa 277--8583, Japan
}
\vskip 2cm
\abstract{

  The recently observed diphoton excess at the LHC may suggest the
  existence of a singlet (pseudo-) scalar particle with a mass of 750
  GeV which couples to gluons and photons. Assuming that the couplings
  to gluons and photons originate from loops of fermions and/or
  scalars charged under the Standard Model gauge groups, we show
  that there is a model-independent upper bound on the cross section
  $\sigma(pp\to S\to \gamma\gamma)$ as a function of the cutoff scale
  $\Lambda$ and masses of the fermions and scalars in the loop.  Such
  a bound comes from the fact that the contribution of each particle
  to the diphoton event amplitude is proportional to its contribution
  to the one-loop $\beta$ functions of the gauge couplings.  We also investigate the perturbativity of
  running Yukawa couplings in models with fermion loops, and show the upper bounds on 
  $\sigma(pp\to S\to \gamma\gamma)$ for explicit models.}

\end{center}
\end{titlepage}

\section{Introduction}
Recently, the ATLAS and CMS collaborations reported an excess of diphoton events implying a resonance with a mass of around $750~\GeV$~\cite{ATLAS-CONF-2015-081,CMS:2015dxe}.
The ATLAS collaboration has 3.2 fb$^{-1}$ of data, and the largest excess is found at around the diphoton invariant mass of $m_{\gamma\gamma}\simeq 750~\GeV$
with the local (global) significance of $3.6\sigma$ ($2.0\sigma$) for a narrow width case.
When a large width for the signal component is assumed, 
the local (global) significance increases to $3.9\sigma$ ($2.3\sigma$) at the width of about $45~\GeV$.
The CMS collaboration, with 2.6 fb$^{-1}$ of data, also reported an excess at around $m_{\gamma\gamma}\simeq 750~\GeV$ with the local (global) significance of $2.6\sigma$ ($1.2\sigma$) 
for a narrow width case, while the significance does not increase with a larger width. 
Possible explanations and implications of this excess have been extensively discussed~\cite{1512.08307, ref:RGEothers, Knapen:2015dap, ref:LoopModels,  ref:OtherStudies}.

One of the plausible explanations of the excess is that a scalar or
pseudoscalar particle $S$ with a mass of $750~\GeV$ is produced
through gluon fusion and decays into a pair of photons, $gg\to S\to
\gamma\gamma$, via diagrams with new fermions and/or bosons charged
under the Standard Model (SM) gauge groups running in the
loops~\cite{1512.08307, ref:RGEothers, Knapen:2015dap, ref:LoopModels}.   In order to
explain the excess with perturbative couplings, however, the new
particles in the loop should have large quantum numbers and/or large
multiplicity, which implies that the perturbativity of the SM gauge
groups may break down at some high scale below the Planck scale.  In
this letter, we address this issue and investigate the perturbativity
of such models.

Our main conclusions are as follows:
\begin{enumerate}
\item We point out that the contribution of each particle in the loop
  to the diphoton event amplitude is proportional to its contribution
  to the one-loop $\beta$ functions of the gauge couplings at the
  leading order, independently of the representations of the particles
  in the loop.  Consequently, there is a generic upper bound on the
  cross section $\sigma(pp\to S\to \gamma\gamma)$ as a function of the
  cutoff scale $\Lambda$ and masses of the fermions and scalars in the
  loop.  We also numerically evaluate such a bound, taking into
  account the following constraints:
\begin{itemize}
\item[(i)] the constraints from Landau pole, requiring that the gauge couplings remain perturbative up to the scale $\Lambda$, and 
\item[(ii)] the constraint from the scale dependence of the strong
  coupling constant based on the LHC~\cite{Becciolini:2014lya}.
\end{itemize}
\item We also investigate the running of the Yukawa coupling in models with fermion loops.  
The upper bound on
  $\sigma(pp\to S\to \gamma\gamma)$ is presented as a function of the
  fermion mass and the cutoff scale $\Lambda$ for some explicit models with 
  vector-like quarks.
\end{enumerate}
The generic analysis in the first part, which can be applied to models
with fermions and scalars in the loop in arbitrary representations,
was not considered in the previous works.  The analysis of the second
part is close to those of Ref.~\cite{1512.08307}, where the authors
investigated the running of the gauge, Yukawa, and scalar quartic
couplings in models with multiple generations of fermions in the
loop. (See also Refs.~\cite{ref:RGEothers} for related works.) They
considered several model points with fixed fermion masses and the
number of generations.  Our analysis is complementary in the sense
that the fermion mass, the Yukawa coupling, the number of generations,
as well as the cutoff scale are taken as free parameters.

In the next section, we investigate the running of gauge couplings in
generic setup with fermions and scalars in arbitrary representations,
and show that there is a model-independent upper bound on the cross
section $\sigma(pp\to S\to \gamma\gamma)$.  In Sec.~\ref{sec:Yukawa},
we investigate the running of the Yukawa coupling (as well as those of gauge couplings) 
in explicit models and present the upper bound on $\sigma(pp\to S\to \gamma\gamma)$ as a
function of the fermion mass and the cutoff scale $\Lambda$. 
We also briefly discuss the LHC constraints on vector-like quarks,
and comment on the running of the
scalar quartic coupling of $S$. We conclude in
Sec.~\ref{sec:conclusion}.

\section{Running gauge couplings and generic upper bound on the
  diphoton event rate}
\label{sec:generic}

The reported diphoton excess can be explained by a new scalar particle
$S$, with a mass of $m_S\simeq 750~\GeV$, which is produced by a gluon
fusion and decays into two photons. The cross section is
given by
\begin{align}
\sigma(pp\to S\to \gamma\gamma)
&=
\frac{C_{gg}}{s\cdot m_S}
\frac{\Gamma(S\to gg)\Gamma(S\to \gamma\gamma)}{
\Gamma_{S,\text{total}}}\,,
\label{eq:ppSgammagamma}
\end{align}
where $\sqrt{s}=13\ {\rm TeV}$ is the center-of-mass energy of the
LHC, and $C_{gg}=(\pi^2/8)\int^1_0 dx_1 \int^1_0 dx_2$
$\delta(x_1x_2-m_S^2/s)g(x_1)g(x_2)$, with $g(x)$ being the gluon
parton distribution function.  In our numerical calculation, we use
the MSTW2008 NLO set~\cite{0901.0002}  evaluated at the
scale $\mu=m_S$, which gives $C_{gg}\simeq 2.1\times 10^3$.  The
reported excess~\cite{ATLAS-CONF-2015-081,CMS:2015dxe} suggests
$\sigma(pp\to S\to \gamma\gamma)\sim {\cal O}(1)$--10~fb.

We assume that the production and the decay of the singlet scalar $S$
is induced through loops of new fermions $\psi_i$ and/or scalars
$\phi_i$. In order to make the analysis model-independent, we consider
that they have generic quantum numbers $(R^{(3)}_i, R^{(2)}_i, Y_i)$
under the SM gauge groups SU(3)$\times$SU(2)$\times$U(1)$_Y$.  The
relevant part of the Lagrangian is given by\footnote{In general,
  off-diagonal couplings such as $y_{ij} S \bar{\psi}_i \psi_j$ and/or
  $A_{ij} S \phi_i^* \phi_j$ $(i\ne j)$ are allowed when $\psi_i$ and
  $\psi_j$ ($\phi_i$ and $\phi_j$) have the same quantum numbers, but
  they do not contribute to the process $gg\to S\to \gamma\gamma$ at the
  one loop.}
\begin{align}
{\cal L}&= {\cal L}_{\text{SM}} 
+ \frac{1}{2}(\partial_\mu S)^2 
- \frac{1}{2}m_S^2 S^2
\nonumber \\
& + \sum_{i=\text{fermions}}
\eta_i \left(
\bar{\psi}_i (i\slashed{D}-m_i) \psi_i 
- y_i S \bar{\psi}_i \psi_i
- iy_{5i} S \bar{\psi}_i \gamma_5 \psi_i 
\right)
\nonumber \\
& + \sum_{i=\text{scalars}}
\eta_i
\left( \left|D_\mu \phi_i\right|^2 - m_i^2 |\phi_i|^2 - A_i S |\phi_i|^2 \right)
+(\text{scalar quartic couplings}),
\end{align}
where $\eta_i=1/2$ for Majorana fermions and real scalars, and
$\eta_i=1$ otherwise.  (Notice that Majorana fermions and real scalars
are possible only for the case of real representation of the SM gauge
group, such as $({\bf 8}, {\bf 1}, 0)$ and $({\bf 1}, {\bf 3}, 0)$.)

In the following, we assume CP-conservation, and consider the two cases of scalar $S$ ($y_{5i}=0$) and pseudoscalar $S$ ($y_i=A_i=0$) separately.
The partial decay rates of $S$ into $gg$ and $\gamma\gamma$ are given by
\begin{align}
\Gamma(S\to gg)
&=
\frac{2}{\pi}\kappa_{gg}^2 m_S^3\,,
\quad
\Gamma(S\to \gamma\gamma)
=
\frac{1}{4\pi}\kappa_{\gamma\gamma}^2 m_S^3\,,
\end{align}
where
\begin{align}
\kappa_{gg} &=
\frac{\alpha_3}{8\pi}\left(
\sum_{i=\text{fermions}} 
\eta_i d^{(2)}_i C^{(3)}_i \frac{y_i}{m_i} \cdot \frac{4}{3}f_{1/2}(\tau_i)
+
\sum_{i=\text{scalars}} 
\eta_i d^{(2)}_i C^{(3)}_i \frac{A_i}{m_i^2} \cdot \frac{1}{6}f_0(\tau_i)
\right)
\,,
\label{eq:Sgg_scalar}
\\
\kappa_{\gamma\gamma} &=
\frac{\alpha_{\text{em}}}{8\pi}\left(
\sum_{i=\text{fermions}}  
\eta_i \text{tr}(Q_i^2) \frac{y_i}{m_i}  \frac{4}{3}f_{1/2}(\tau_i)
+
\sum_{i=\text{scalars}} 
\eta_i \text{tr}(Q_i^2) \frac{A_i}{m_i^2} \cdot \frac{1}{6}f_0(\tau_i)
\right)\,,
\end{align}
in the case of scalar $S$, and 
\begin{align}
\kappa_{gg} &=
\frac{\alpha_3}{8\pi}
\sum_{i=\text{fermions}} 
\eta_i d^{(2)}_i C^{(3)}_i \frac{y_{5i}}{m_i} \cdot 2\widetilde{f}_{1/2}(\tau_i)\,,
\\
\kappa_{\gamma\gamma} &=
\frac{\alpha_{\text{em}}}{8\pi}
\sum_{i=\text{fermions}} 
\eta_i \text{tr}(Q_i^2) \frac{y_{5i}}{m_i} \cdot 2\widetilde{f}_{1/2}(\tau_i)\,,
\label{eq:Sgmmgmm_pseudoscalar}
\end{align}
in the case of pseudoscalar $S$.  Here, $d_i^{(N)}$ and $C^{(N)}_i$
are the dimension and the Dynkin index of the representation
$R_i^{(N)}$ of SU$(N)$, respectively.  For instance, $(d_i^{(N)},
C_i^{(N)})=(1,0)$, $(N,1/2)$ and $(N^2-1,N)$ for $R_i^{(N)}$ being
singlet, fundamental representation, and adjoint representation,
respectively.  The trace of the electric charge squared is given by
\begin{align}
\text{tr}(Q_i^2)=d_i^{(3)}C_i^{(2)}+d_i^{(3)}d_i^{(2)} Y_i^2\,,
\end{align}
and the loop functions are (for $\tau<1$)
\begin{align}
f_{1/2}(\tau) &=\frac{3}{2\tau^2}\left(\tau+(\tau-1)\arcsin^2\sqrt{\tau}\right)\,,
\\
\widetilde{f}_{1/2}(\tau) &= \frac{1}{\tau} \arcsin^2\sqrt{\tau}\,,
\\
f_0(\tau) &=\frac{3}{\tau^2}\left( \arcsin^2\sqrt{\tau}  - \tau\right)\,,
\end{align}
with $\tau_i=m_S^2/4m_i^2$.  (These loop functions are normalized so
that they become $1$ for $\tau\to 0$.)

Let us now discuss the running gauge couplings of the SM for a scale
at $\mu>m_i$, which are give by, at the one loop,
\begin{align}
\alpha_a^{-1}(\mu) \simeq \alpha_{a,\text{SM}}^{-1}(m_i) 
- \frac{b_a^{\text{SM}}+\Delta b_a}{2\pi}\log\lrf{\mu}{m_i}\,,
\label{eq:running}
\end{align}
where $\alpha_{a,\text{SM}}(m_i)$ is evaluated by using the
renormalization group (RG) equations of the SM,
$b_a^{\text{SM}}=(41/6, -19/6, -7)$, and
\begin{align}
\Delta b_a
&=
\frac{4}{3}
\sum_{i=\text{fermions}}
\eta_i
\begin{pmatrix}
d^{(3)}_i d^{(2)}_i Y_i^2
\\
d^{(3)}_i C^{(2)}_i
\\
d^{(2)}_i C^{(3)}_i
\end{pmatrix}
+
\frac{1}{3}
\sum_{i=\text{scalars}}
\eta_i
\begin{pmatrix}
d^{(3)}_i d^{(2)}_i Y_i^2
\\
d^{(3)}_i C^{(2)}_i
\\
d^{(2)}_i C^{(3)}_i
\end{pmatrix}.
\label{eq:Deltab}
\end{align}
Note that the contributions of each fermion or
scalar to $\Delta b_a$ in \eqref{eq:Deltab} are the same
as the coefficients in the diphoton production rate,
Eqs.~\eqref{eq:Sgg_scalar}-\eqref{eq:Sgmmgmm_pseudoscalar}.  
Therefore, by defining effective masses $m_i^{\text{eff}}$ and its minimal value as
\begin{align}
m_i^{\text{eff}} &\equiv 
\begin{cases}
\displaystyle{\frac{m_i}{y_i f_{1/2}(\tau_i)}} 
\;\text{or}\;
\displaystyle{\frac{m_i}{y_{5i} \tilde{f}_{1/2}(\tau_i)}} 
& (i=\text{fermion})\,,
\\
\displaystyle{\frac{2m_i^2}{A_i f_0(\tau_i)}} & (i=\text{scalar})\,,
\end{cases}
\nonumber\\
m_{\text{min}}^{\text{eff}} &\equiv \min_i\{m_i^{\text{eff}}\}\,,
\label{eq:meff}
\end{align}
one can obtain upper bounds on $\Gamma(S\to gg)$ and $\Gamma(S\to
\gamma\gamma)$ as functions of $\Delta b_a$ and
$m_{\text{min}}^{\text{eff}}$ as follows;
\begin{align}
\Gamma(S\to gg) < 
\Gamma(S\to gg)_{\max}
&=
\frac{2m_S^3}{\pi}
\lrfp{\alpha_3}{8\pi}{2}
\lrfp{\Delta b_3}{m_{\text{min}}^{\text{eff}}}{2}
\times\begin{cases}
1 & (\text{scalar}~S)
\\
9/4 & (\text{pseudoscalar}~S)
\end{cases},
\label{eq:GammaMaxGluon}
\\
\Gamma(S\to \gamma\gamma)
<
\Gamma(S\to \gamma\gamma)_{\max}
&=
\frac{m_S^3}{4\pi}
\lrfp{\alpha_{\text{em}}}{8\pi}{2}
\lrfp{\Delta b_1 + \Delta b_2}{m_{\text{min}}^{\text{eff}}}{2}
\times\begin{cases}
1 & (\text{scalar}~S)
\\
9/4 & (\text{pseudoscalar}~S)
\end{cases}.
\label{eq:GammaMaxPhoton}
\end{align}

We consider the following two constraints on $\Delta b_a$:
\begin{itemize}
\item[(i)] {\bf Landau pole}: We require that the SM gauge couplings
  are perturbative up to a scale $\Lambda$,\footnote{We have checked
    that the numerical results are almost unchanged as far as
    $\alpha_a(\Lambda)$ is larger than $1$.}
\begin{align}
\alpha_a(\Lambda) < 1\,,
\end{align}
which leads to upper bounds on $\Delta b_a$ as functions of $\Lambda$
(and $m_i$).
\item[(ii)] {\bf Running $\alpha_3$}: In addition, too large $\Delta
  b_3$ (with relatively small $m_i$) modifies the evolution of the
  strong coupling constant and conflicts with the scale dependence of $\alpha_3$
  observed by the LHC~\cite{Becciolini:2014lya}.  We require that
  $\Delta b_3$ is below the $2\sigma$ upper bound given in
  Ref.~\cite{Becciolini:2014lya}.\footnote{Although there are also
    similar bounds on $\Delta b_{1,2}$ from the measurements of
    running electroweak couplings $\alpha_{1,2}$~\cite{Alves:2014cda},
    we found that the constraints are too weak to constrain the
    diphoton models.}  For instance, the bound is $\Delta b_3<5.2$
  (15.9) when the mass of the particle in the loop is 500 (700) GeV.
\end{itemize}
These bounds on $\Delta b_a$ lead to the maximal values of
$\Gamma(S\to gg)$ and $\Gamma(S\to \gamma\gamma)$ according to
Eqs.~\eqref{eq:GammaMaxGluon} and \eqref{eq:GammaMaxPhoton}, which are
then converted to the upper bound on the cross section for the process
$pp\rightarrow S\rightarrow\gamma\gamma$.  In particular, as one can
see from Eq.\ \eqref{eq:ppSgammagamma}, the cross section becomes
larger as $\Gamma(S\to gg)$ increases.  In addition, when $\Gamma(S\to
gg)$ takes its largest possible value, the partial decay rates into
electroweak gauge boson pairs are always much smaller than
$\Gamma(S\to gg)$, and $\sigma (pp\rightarrow
S\rightarrow\gamma\gamma)$ increases as $\Gamma(S\to \gamma\gamma)$
becomes larger.  Thus, with $\Lambda$ and
$m_{\text{min}}^{\text{eff}}$ being fixed, the cross section takes its
largest value when $\Delta b_1$, $\Delta b_2$ and $\Delta b_3$ are all
maximized.

\begin{figure}[t]
\begin{center}
\includegraphics[width=8cm]{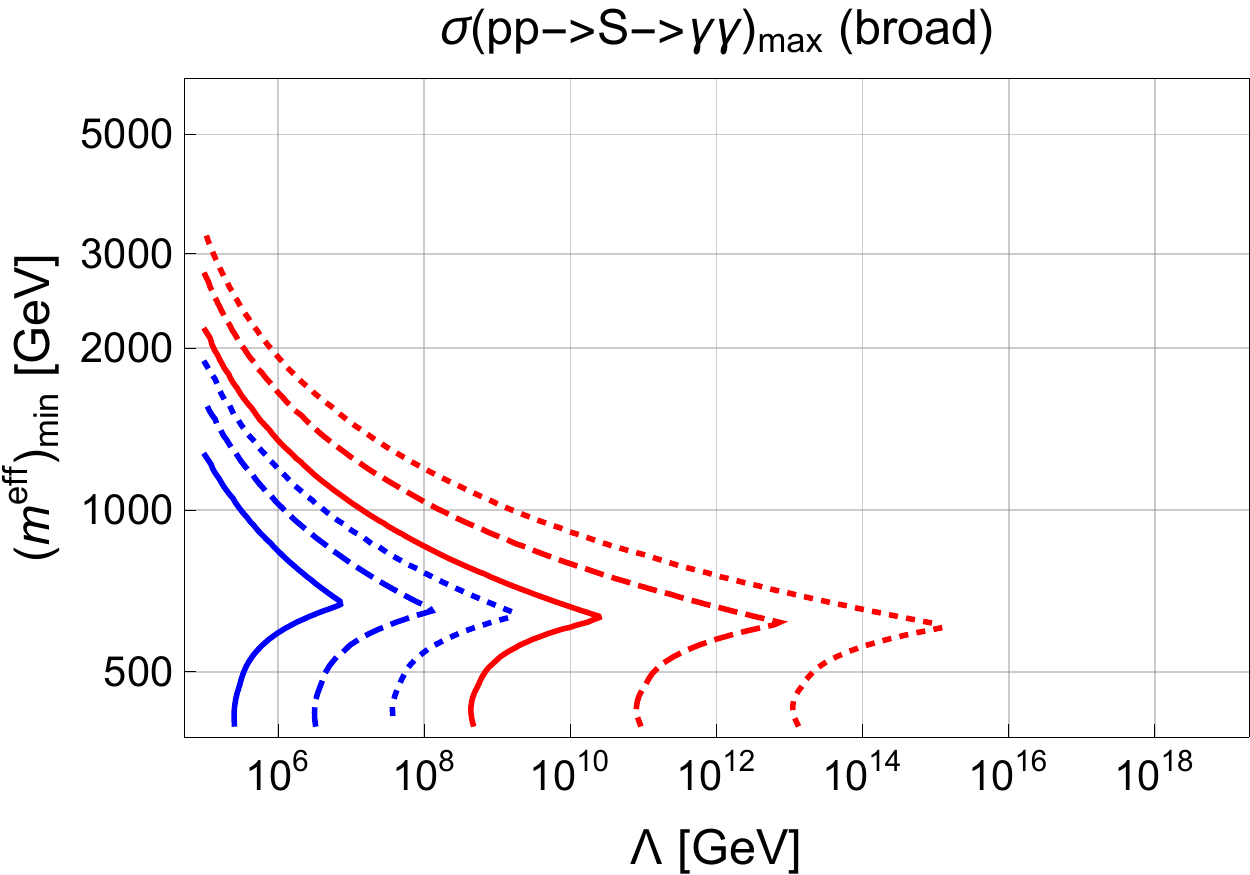}
\includegraphics[width=8cm]{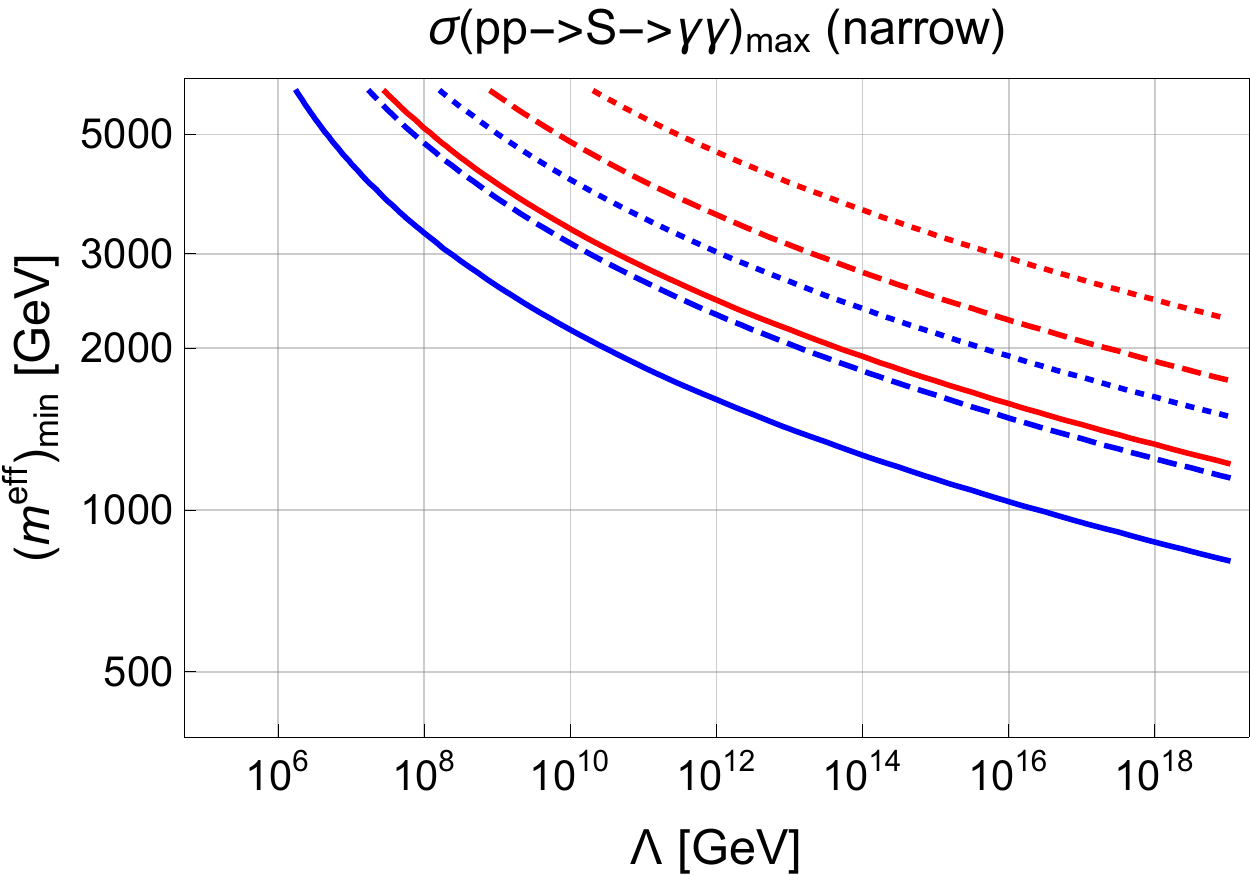}
\end{center}
\caption{ Contour plots of the upper bounds on the signal rate
  $\sigma(pp\to S\to \gamma\gamma)_{\max}$ as functions of the cutoff
  scale $\Lambda$ and the minimum effective mass in the loop,
  $m^{\text{eff}}_{\text{min}}$, defined in Eq.~\eqref{eq:meff}.
  Left: broad width case, $\Gamma_{S,\text{total}}=45~\GeV$.  Right:
  narrow width case, $\Gamma_{S,\text{total}}=\Gamma(S\to
  gg)+\Gamma(S\to \gamma\gamma)$.  The solid, dashed, and dotted lines
  show the contours of $\sigma(pp\to S\to
  \gamma\gamma)_{\max}=10$, $5$, and
  $3~\text{fb}$, respectively.  The blue lines represent the case that
  $S$ is a scalar, while red lines are for the pseudoscalar case.  }
\label{fig:maxGammaAndDiphoton}
\end{figure}

Fig.~\ref{fig:maxGammaAndDiphoton} shows the upper bound on
$\sigma(pp\to S\to \gamma\gamma)$ as a function of $\Lambda$ and
$m^{\text{eff}}_{\text{min}}$, which is obtained from
Eqs.~\eqref{eq:ppSgammagamma}, \eqref{eq:GammaMaxGluon}, and
\eqref{eq:GammaMaxPhoton}.  The left figure shows the case of fixed
broad width $\Gamma_{S,\text{total}}=45~\GeV$, while the right figure
represents the case of narrow width,
$\Gamma_{S,\text{total}}=\Gamma(S\to gg)+\Gamma(S\to
\gamma\gamma)$.\footnote
{We have checked that the result does not change much even if we include
  other decay modes into electroweak gauge boson pairs.}
The red and blue lines show the cases that $S$ is a scalar and a
pseudoscalar, respectively.  Here, for simplicity, we have taken
$m_i=m^{\text{eff}}_{\text{min}}$ to calculate the running coupling
with Eq.~\eqref{eq:running}, and also to obtain the upper bound on
$\Delta b_3$ from Ref.~\cite{Becciolini:2014lya}.\footnote{As we shall
  see in the next section, the physical masses $m_i$ are typically
  smaller than the effective mass $m^{\text{eff}}_i$ in concrete
  models. Smaller masses give severer upper bounds on $\Delta b_a$ for
  both of the constraints (i) and (ii), and hence taking
  $m_i=m^{\text{eff}}_{\text{min}}$ leads to conservative
  constraints.}

As can be seen in the left panel of
Fig.~\ref{fig:maxGammaAndDiphoton}, the cutoff scale $\Lambda$ cannot
be very large for a broad width case. Below the kink at $m\simeq
600$--$700$ GeV, the constraint from the $\alpha_3(\mu)$ measurement
gives a severe upper bound on $\Delta b_3$. In this region, the upper
bound on $\Lambda$ is determined by the condition of Landau poles of
$\alpha_{1,2}$. Above the kink, the Landau pole condition on $\Delta
b_3$ is stronger than that from the $\alpha_3(\mu)$ measurement.

In the narrow width case shown in the right panel of Fig.~\ref{fig:maxGammaAndDiphoton}, the bounds become weaker than the broad width case, but they still constrain the region of $m^{\text{eff}}_{\text{min}}\simeq {\cal O}(\TeV)$ when the cutoff scale $\Lambda$ is large.
For instance, in order to have
$\sigma(pp\to S\to \gamma\gamma)=10$~fb with $\Lambda=10^{18}~\GeV$
($10^{15}~\GeV$), the effective mass should be
$m^{\text{eff}}_{\text{min}}\lesssim 870$ (1100)~GeV in the case that $S$ is a scalar, 
and 
$m^{\text{eff}}_{\text{min}}\lesssim 1300$ (1700)~GeV in the case of pseudoscalar.

Before closing this section, several comments are in order. 
\begin{itemize}
\item The bounds in Fig.~\ref{fig:maxGammaAndDiphoton} are very
  conservative, and they can become severer in concrete and realistic
  models. First of all, $\Delta b_1$, $\Delta b_2$ and $\Delta b_3$
  are simultaneously maximized in Fig.~\ref{fig:maxGammaAndDiphoton},
  but it is not generically the case in concrete models.  Secondly,
  the region with small mass and large $\Delta b_a$ is severely
  constrained by the direct search for the new particles $\psi_i$ and
  $\phi_i$. For instance, in the broad width case, the constraint from
  the $\alpha_3(\mu)$ measurement for $m=600~\GeV$ is about $\Delta
  b_3 < 8.7$, and the upper bound corresponds to 13 Dirac pairs of
  vector-like quarks if they are in fundamental representations. Such
  a model is likely to be already excluded by direct searches, unless
  the new colored particles decay in a very complicated manner to escape from LHC
  searches. The direct search can constrain the model for the narrow
  width case as well. (See also the discussion in the next section.)

Although it is difficult to saturate the bounds in Fig.~\ref{fig:maxGammaAndDiphoton} in concrete realistic models, they are model-independent and conservative, and yet constraining interesting regions of $m^{\text{eff}}_{\text{min}}$ and $\Lambda$. Therefore the bounds in Fig.~\ref{fig:maxGammaAndDiphoton} can be an important first step to explore the physics behind the diphoton signal.

\item In models with fermion loops, the Yukawa coupling 
  $y_{(5)i}$ at low
  energy becomes typically smaller than unity due to the running, and
  hence the masses of the particles in the loop $m_i$ should be even
  smaller than $m^{\text{eff}}_{\text{min}}$
  (cf. Eq.~(\ref{eq:meff})).  In other words, if one adjusts the
  Yukawa couplings at TeV scale to larger values, the scale of the
  Landau pole of the Yukawa coupling becomes even smaller than those
  of the gauge couplings.  (See the next section.)
\end{itemize}

\section{Explicit examples}
\label{sec:Yukawa}

In the previous section, we have derived a generic upper bound on
$\sigma(pp\to S\to \gamma\gamma)$ for given cutoff scale $\Lambda$ and
effective mass scale $m^{\rm eff}_{\rm min}$.  Although it is a
prominent implication of the diphoton resonance, $m^{\rm eff}_{\rm
  min}$ does not directly correspond to physical masses of new charged
particles.  In order to see how light the charged particles should be
in models with fermion loops, in this section we consider the running
of the Yukawa couplings with concrete examples.  We also briefly
discuss the LHC constraints on vector-like quarks, and comment on the
running of the scalar quartic coupling of $S$.  In this section, we
only consider the case of narrow width and take
$\Gamma_{S,\text{total}}=\sum_{VV=gg,\gamma\gamma,\gamma
  Z,ZZ,WW}\Gamma(S\to VV)$.

For simplicity, we consider the $N$ copies of Dirac fermions which transform
as $(R^{(3)},R^{(2)},Y)$ under the SM gauge group, with universal
Yukawa coupling and mass, $y_{(5)i}=y$ and $m_i=m$.  The RG equation for
Yukawa coupling is given by~\cite{gen_rge}
\begin{equation}
16\pi^2\frac{dy}{d\ln\mu}=\left(3+2d^{(2)}d^{(3)}N \right)y^3-
\left( 
\frac{48C^{(3)}}{d^{(3)}} g_3^2 + 
\frac{18C^{(2)}}{d^{(2)}} g_2^2 + 
6Y^2 g_1^2
\right)
y\,,
\label{eq:simple_case}
\end{equation}
which holds both for scalar $S$ ($y=y_i$) and pseudoscalar $S$
($y=y_{5i}$).  
For a given
representation $(R^{(3)},R^{(2)},Y)$, one can obtain the upper bound on
$\sigma(pp\to S\to\gamma\gamma)$ as a function of $m$ and $\Lambda$
from the following procedure.
\begin{enumerate}
\item An upper bound on the multiplicity $N$, $N_{\text{max}}$, is
  obtained as a function of $m$ and $\Lambda$, by requiring that (i)
  the gauge couplings remain perturbative up to the scale $\Lambda$,
  and (ii) $\Delta b_3$ satisfies the constraint from the
  $\alpha_3(\mu)$ measurement~\cite{Becciolini:2014lya} (see
  Sec.~\ref{sec:generic}).  For the former constraint, we require
  $\alpha_a(\Lambda)\leq1$ for $a=1$--$3$ in the numerical calculation.
  
\item For a given $N$ ($1\le N\le N_{\text{max}}$), an upper bound on
  the Yukawa coupling at low energy, $y(\mu = m)$, is obtained by
  requiring that the running Yukawa coupling, $y(\mu)$, also remains
  perturbative for $\mu< \Lambda$.  This gives the upper bounds on
  $\sigma(pp\to S\to\gamma\gamma)$ for a given set of
  $(m,\Lambda,N)$.  Because $y(m)$ increases as $y(\Lambda)$ increases,
  we take $y(\Lambda)=4$.  (We have checked that the maximal possible
  value of the cross section does not change much as far as
  $y(\Lambda)$ is large enough.)
\item The maximum signal rate $\sigma(pp\to
  S\to\gamma\gamma)_{\text{max}}$ is obtained with respect to $N$.
\end{enumerate}
In the case where there is only one representation, $N=N_{\rm max}$
gives the maximum value of $\sigma(pp\to S\to \gamma\gamma)$ with $m$,
$\Lambda$, and the representation of the fermion being
fixed. This is because the maximal value of the Yukawa coupling at low energy roughly scales as $y\sim N^{-1/2}$, 
and therefore the signal rate increases as $\sigma(pp\to S\to \gamma\gamma)\sim (N y)^2\sim N$.

As explicit examples, we further assume that the new vector-like fermions can
decay into SM particles with a renormalizable interaction.  Then, there are
seven possibilities~\cite{Knapen:2015dap}
\begin{align}
(R^{(3)},R^{(2)},Y) =&
(\mathbf{3},\mathbf{1},-1/3), (\mathbf{3},\mathbf{1},2/3), 
\nonumber\\
& (\mathbf{3},\mathbf{2},1/6), (\mathbf{3},\mathbf{2},-5/6), (\mathbf{3},\mathbf{2},7/6), 
\nonumber \\
& (\mathbf{3},\mathbf{3},-1/3), (\mathbf{3},\mathbf{3},2/3).
\label{eq:reps}
\end{align}
For those representations, we have numerically solved the RG equation in
\eqref{eq:simple_case} as well as those of gauge coupling constants,
and calculated the maximum signal rate $\sigma(pp\to S\to\gamma\gamma)_{\text{max}}$.
In the following, we mainly discuss three cases
$(\mathbf{3},\mathbf{1},2/3)$, $(\mathbf{3},\mathbf{2},7/6)$, and $(\mathbf{3},\mathbf{3},2/3)$,
since they have large hyper-charges and give largest signal rate among
SU(2) singlets, doublets and triplets, respectively.
We will briefly discuss the other cases at the end of this section.

\begin{figure}[t!]
\begin{center}
\includegraphics[width=8cm]{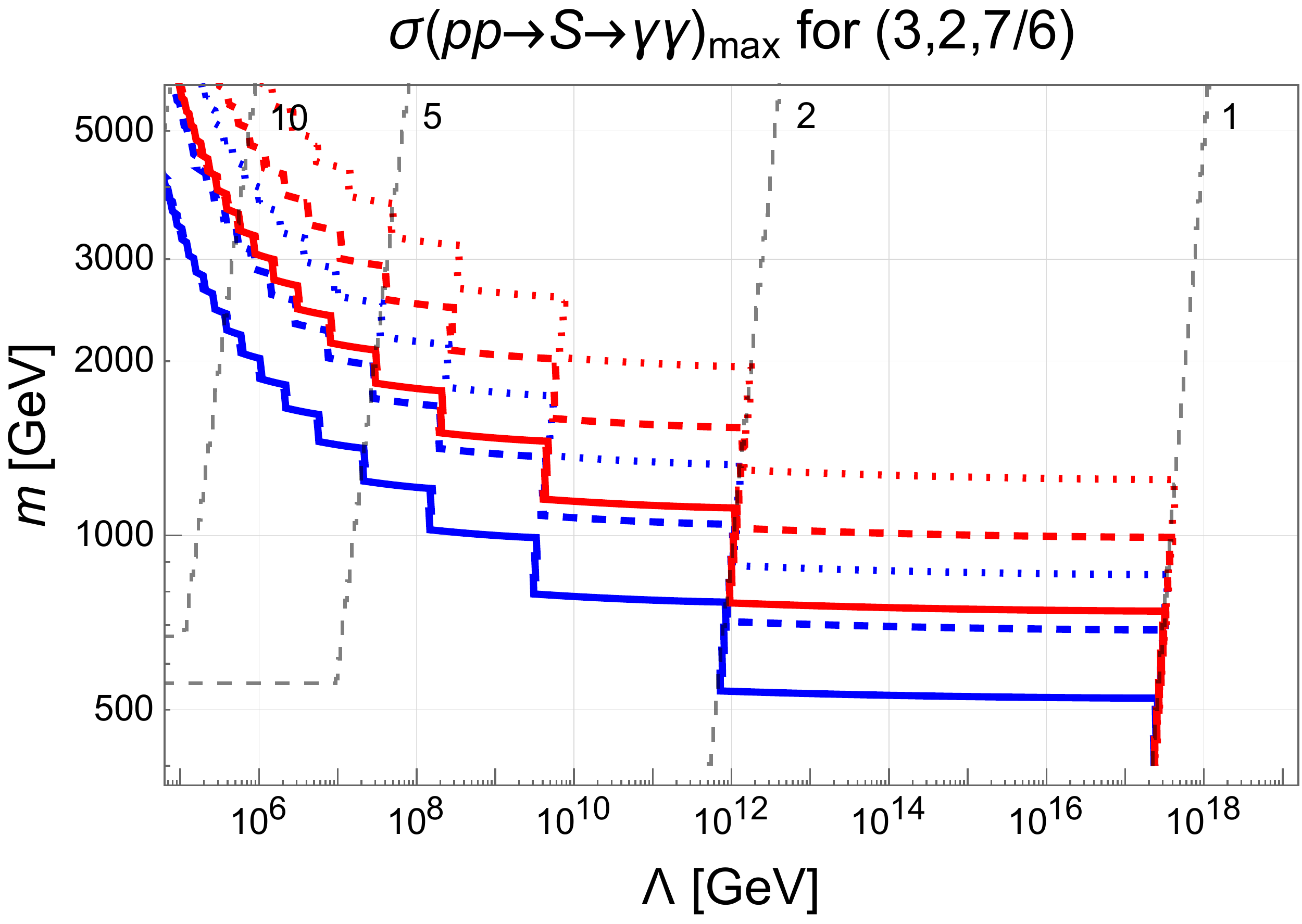}
\includegraphics[width=8cm]{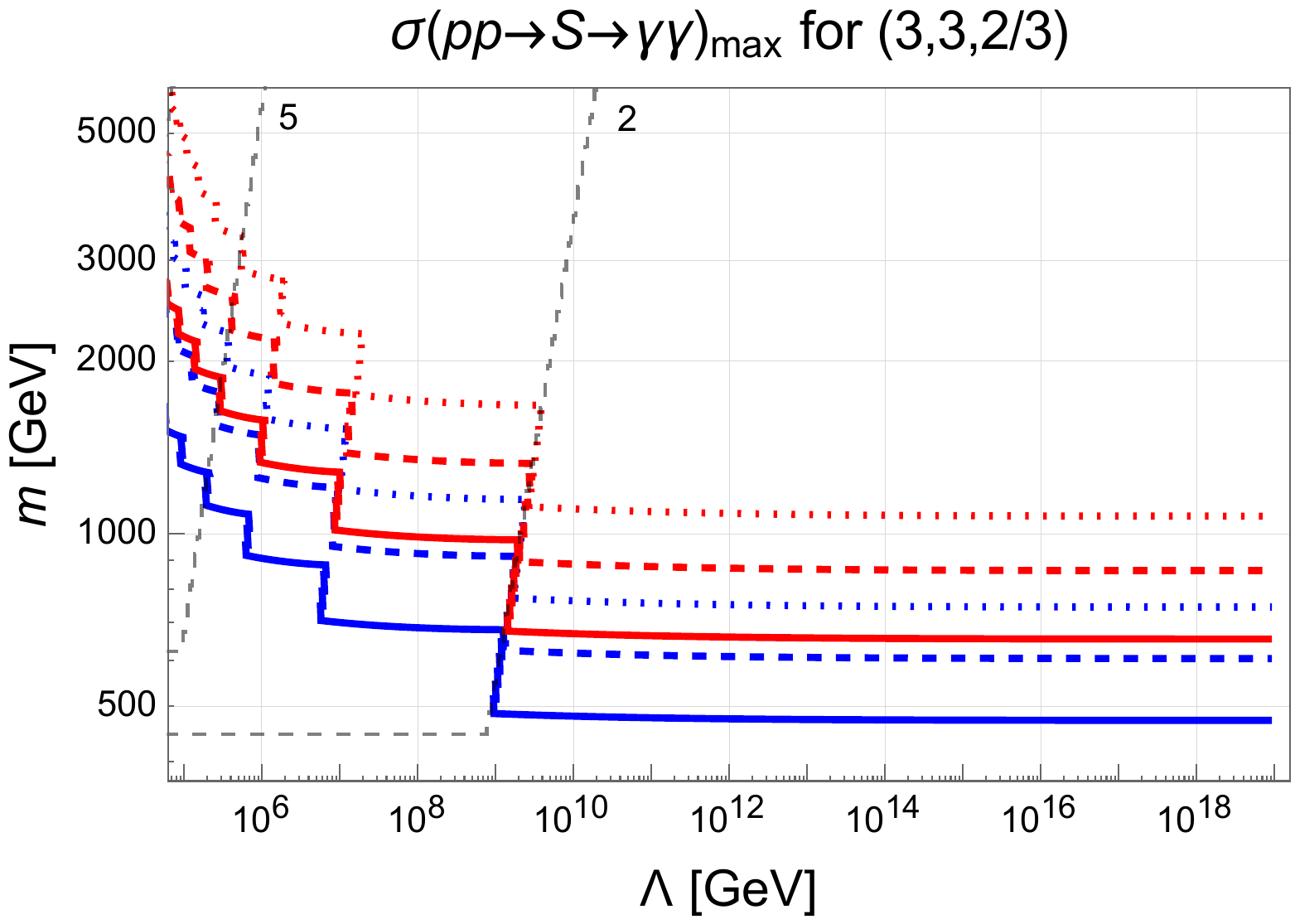}
\\
\includegraphics[width=8cm]{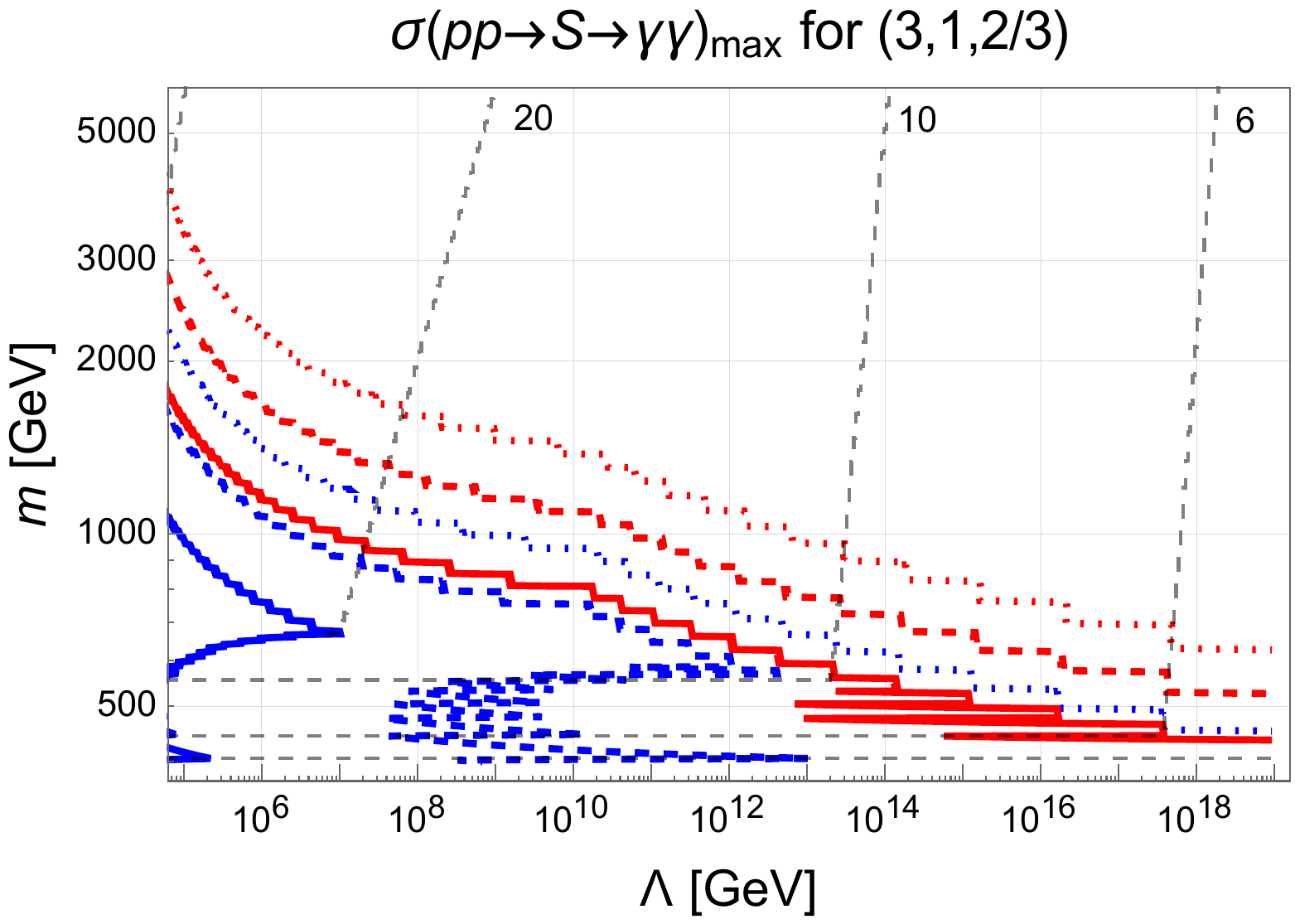}
\end{center}
\caption{ The upper bound on the signal rate $\sigma(pp\to S\to
  \gamma\gamma)_{\max}$ as a function of the cutoff scale $\Lambda$
  and the fermion mass $m$ in the case of Dirac fermions with
  quantum numbers of 
  $(\mathbf{3},\mathbf{2},7/6)$ (top left), 
  $(\mathbf{3},\mathbf{1},2/3)$ (bottom), and
  $(\mathbf{3},\mathbf{3},2/3)$ (top right).
Here, we take $\Gamma_{S,\text{total}}=\sum_{VV=gg,\gamma\gamma,\gamma
  Z,ZZ,WW}\Gamma(S\to VV)$.
The solid, dashed, and dotted lines show the contours of $\sigma(pp\to S\to
  \gamma\gamma)_{\max}=10$, $5$, and
  $3~\text{fb}$, respectively.  The blue lines represent the case that
  $S$ is a scalar, while red lines are for the pseudoscalar
  case. Black dashed lines show the maximal allowed number of
  generations, $N_{\text{max}}$. }
\label{fig:sigma_max}
\end{figure}

The results for
$(\mathbf{3},\mathbf{1},2/3)$, $(\mathbf{3},\mathbf{2},7/6)$, and $(\mathbf{3},\mathbf{3},2/3)$
are shown in Fig.~\ref{fig:sigma_max}.
As one can see, they look qualitatively  similar to
Fig.\ \ref{fig:maxGammaAndDiphoton}.  However, with
$m^{\text{eff}}_{\rm min}$ being fixed, the cross section is smaller
than that given in Fig.\ \ref{fig:maxGammaAndDiphoton} because $\Delta
b_1$ and $\Delta b_2$ are not simultaneously maximized.  Furthermore,
$m$ is smaller than $m^{\text{eff}}_{\rm min}$ because the Yukawa coupling
becomes smaller than unity at low energy, in particular when $N$ is
large (cf. Eq.~\eqref{eq:simple_case}).
\begin{itemize}
\item
Among the three cases, $(\mathbf{3},\mathbf{2},7/6)$ gives the largest 
$\sigma(pp\to S\to \gamma\gamma)_{\text{max}}$ in most of the parameter space.
For instance, in the pseudoscalar case, $\sigma(pp\to S\to \gamma\gamma)=10$ fb can be realized 
with a cutoff scale of $10^{16}~\GeV$ ($10^{10}~\GeV$)
if the fermions are lighter than 740 GeV (1100 GeV).
For $\sigma(pp\to S\to \gamma\gamma)=5$ fb, the fermions can be as heavy as 1000 GeV (1600 GeV) for  the cutoff scale of $10^{16}~\GeV$ ($10^{10}~\GeV$).
\item
In the case of SU(2) triplet $(\mathbf{3},\mathbf{3},2/3)$, the maximal signal rate 
$\sigma(pp\to S\to \gamma\gamma)_{\text{max}}$ is smaller than the case of $(\mathbf{3},\mathbf{2},7/6)$
in most of the parameter space, except for the large cutoff region $\Lambda\gtrsim 10^{17}\GeV$.
In this case, the cutoff above the Planck scale is allowed, e.g., 
for the pseudoscalar case with $N=1$, $m\simeq 860~\GeV$, and 
$\sigma(pp\to S\to \gamma\gamma)=5$~fb.
\item
Finally, in the case of SU(2) singlet $(\mathbf{3},\mathbf{1},2/3)$, 
the signal rate is suppressed compared with the other two cases.
In this case, the running $\alpha_3$ constraint determines the $N_{\text{max}}$
in a large part of the low mass region $m\lesssim 600$--700 GeV.
The zigzag lines for $m\lesssim 600$ GeV is due to 
the rapid increase of allowed $N_{\text{max}}$ ($\Delta b_3$) with respect to $m$
from the running $\alpha_3$ constraint. In each narrow range of $m$ with a fixed $N_{\text{max}}$,
the upper bound on $\Lambda$ is determined either by the 
perturbativity of the Yukawa coupling or by the Landau pole of U(1)$_Y$.
\end{itemize}

Next, we discuss the constraint from the direct searches for
vector-like quarks at the LHC. Here, we assume that they decay into
the SM particles via a renormalizable coupling with SM quarks and the
Higgs boson. In order to avoid the stringent constraint from the decay
into third generation quarks, let us further assume that the coupling
with the third generation is suppressed. Then, the vector-like quarks
decay into a light SM quark and a $W$/$Z$/Higgs boson, depending on
its representation.  In particular, the search for a vector-like quark
decaying into a $W$ boson and a light SM quark at the LHC gives a
stringent constraint in the present scenario.  From the result of
ATLAS~\cite{Aad:2015tba}, the bound is estimated as\footnote{We could
  not find any constraint for vector-like quarks with $m>800\ \GeV$
  decaying into light quarks.  Thus, we do not consider the direct bound
  for  $m>800\ {\rm GeV}$.}
\begin{align}
\begin{cases}
N\cdot \text{Br}(Q'\to Wq) \lesssim 1 & ~\text{for}~m \le 690~\GeV\,,
\\
N\cdot \text{Br}(Q'\to Wq) \lesssim 1.5 & ~\text{for}~690~\GeV \le m \lesssim 750~\GeV\,,
\\
N\cdot \text{Br}(Q'\to Wq) \lesssim 2 & ~\text{for}~750~\GeV \lesssim m \lesssim 800~\GeV\,,
\end{cases}
\label{eq:Wq_bound}
\end{align}
where $Q'$ and $q$ denote the vector-like quark and the SM light quark, respectively.
\begin{itemize}
\item In the case of the SU(2) doublet $(\mathbf{3},\mathbf{2},7/6)$,
  it contains vector-like quarks with electric charges of $5/3$ and
  $2/3$. The one with the electric charge of $5/3$ decays into a $W$
  boson and a light SM quark (up and/or charm) with almost 100\%
  branching fraction.  Comparing the bound in \eqref{eq:Wq_bound} with
  the lines in Fig.~\ref{fig:sigma_max}, if we require $\sigma(pp\to
  S\to \gamma\gamma)=10$ fb in the case of scalar $S$, the region of
  $\Lambda\gtrsim 10^{12}~\GeV$ ($N=1$) is excluded, and
  $\Lambda\gtrsim 10^{9.5}~\GeV$ ($N=2$) is at the boundary of
  excluded region.  In the case of pseudoscalar, the model can explain
  $\sigma(pp\to S\to \gamma\gamma)=10$ fb while being perturbative up
  to $\Lambda\simeq 10^{17}~\GeV$ ($\Lambda\simeq 10^{12}~\GeV$), if
  the vector-like quarks are as light as about 740 GeV (1100 GeV).

\item
In the case of SU(2) triplet $(\mathbf{3},\mathbf{3},2/3)$, 
one of the SU(2) triplet quarks decays into $Wq$ with an almost 100\% branching fraction,
and another one has about 50\% branching. 
Thus, from the bound \eqref{eq:Wq_bound}, the region of $m\lesssim 750~\GeV$ is excluded even for $N=1$. The scalar case cannot have a cutoff larger than about $10^9~\GeV$
in order to have the cross section larger than $\sim 3\ {\rm fb}$, 
while the pseudoscalar case with 
$N=1$, $m\simeq 860~\GeV$, and 
$\sigma(pp\to S\to \gamma\gamma)=5$ fb
is still allowed and can be perturbative up to the Planck scale.

\item Finally, in the case of SU(2) singlet
  $(\mathbf{3},\mathbf{1},2/3)$, the direct search excludes a large
  fraction of the parameter space with a sizable signal cross section,
  in particular when the cutoff scale is high.  In this case, the
  vector-like quark decays into a $W$ boson and a light quark with a
  branching fraction of about 50\%.  From the direct search bound in
  Eq.~\eqref{eq:Wq_bound}, the number of multiplicity should satisfy
  $N<2$, 3, and 4 for $m\le 690~\GeV$, $m\simeq (690$--$750)~\GeV$,
  and $m\simeq (750$--$800)~\GeV$, respectively.  Thus, the lines in
  the figure for $m\lesssim 800~\GeV$ are not consistent with the
  direct search bound.  If we adopt the maximal number of multiplicity
  allowed by the direct search, the Yukawa coupling should be quite
  large at low energy in order to explain the diphoton signal.  Even
  for the pseudoscalar case and for $\sigma(pp\to S\to
  \gamma\gamma)=3$ fb, the required value of the Yukawa coupling is
  $y(m) \gtrsim 2.1$, $1.5$, and $1.2$, for $m\le 690~\GeV$, $m\simeq
  (690$--$750)~\GeV$, and $m\simeq (750$--$800)~\GeV$, respectively.
  If the RG equation \eqref{eq:simple_case} is evolved from low energy
  to high energy, they quickly become non-perturbative, which leads to
  cutoff scales below 10 TeV.
\end{itemize}
We should note that the above constraints strongly depend on the decay modes of vector-like quarks. If they mainly couple to the third generation SM quarks and decay into top and/or bottom quarks, the constraints become severer. 
Instead, if they decay in a very complicated way (e.g., in a cascade decay chain with multiple intermediate new particles emitting many soft jets), they may escape the direct search even for small mass region.

Now let us briefly discuss the other representations in Eq.~\eqref{eq:reps}.
\begin{itemize}
\item In the case of $(\mathbf{3},\mathbf{1},-1/3)$, we checked that,
  even for the pseudoscalar case, $\sigma(pp\to S\to \gamma\gamma)$ is
  smaller than $3$ fb for $\Lambda>10^5~\GeV$.
\item
In the case of $(\mathbf{3},\mathbf{2},1/6)$, 
the pseudoscalar case can have $\sigma(pp\to S\to \gamma\gamma)=(5$--10) fb
with a large cutoff, but it requires a small mass $m$ and a  large multiplicity $N$.
We found that the region below $m\lesssim 800~\GeV$ is excluded if
the vector-like quarks mainly decay into the SM light quarks, 
and for $m\gtrsim 800~\GeV$ the cutoff cannot be larger than $10^9\GeV$
for $\sigma(pp\to S\to \gamma\gamma)\ge 5$ fb.
\item The case of $(\mathbf{3},\mathbf{2},-5/6)$ is similar to that of
  $(\mathbf{3},\mathbf{2},7/6)$, but with the multiplicity $N$ roughly
  twice as large.  For instance, $\sigma(pp\to S\to
  \gamma\gamma)\simeq 5$ fb can be obtained in the pseudoscalar case
  with $N=2$, $m\simeq 910~\GeV$, and the cutoff scale as large as
  $\Lambda\simeq 10^{17}~\GeV$.
\item
The case of $(\mathbf{3},\mathbf{3},-1/3)$ is similar to $(\mathbf{3},\mathbf{3},2/3)$, 
but with the fermion mass $m$ being about 30\% smaller.
\end{itemize}

Before closing this section, we comment on the 
running of the quartic coupling of the $S$ field.\footnote{We assume that
there is no direct coupling between $S$ and the SM Higgs.}
Defining the coupling $\lambda$ as ${\cal L}_{S^4} = -(1/4!)\lambda S^4$, 
its RG equation is given by
\begin{align}
16\pi^2\frac{d\lambda}{d\ln\mu} 
&= 3\lambda^2  -48Nd^{(3)}d^{(2)}y^4 + 8Nd^{(3)}d^{(2)}y^2 \lambda\,.
\end{align}
We have checked that, as far as $\lambda$ is positive at the cutoff
scale, it does not become negative for $m<\mu<\Lambda$ and hence there
is no vacuum instability.  In addition, $\lambda$ does not blow up
below the cutoff scale irrespective of the value $\lambda
(\Lambda)$. Thus, there is no constraint from the running of the
quartic coupling.

\section{Conclusions}
\label{sec:conclusion}

Motivated by the recent LHC results, we have studied the diphoton
resonance production cross section at the LHC, paying particular
attention to the running of the gauge and Yukawa coupling constants.
We have considered the case where a (pseudo-)scalar particle $S$ with its mass
of $750\ {\rm GeV}$ is responsible for the diphoton events observed by
the LHC and the scalar particle is produced by the gluon fusion.  In
such a case, new fermions and/or bosons which have SM gauge quantum
numbers are necessary to generate $S$-$g$-$g$ and
$S$-$\gamma$-$\gamma$ vertices.  Assuming that the $S$-$g$-$g$ and
$S$-$\gamma$-$\gamma$ vertices are perturbatively generated by the
loop effects of the new fermions and/or bosons, we studied how large
the cross section for the process $pp\rightarrow
S\rightarrow\gamma\gamma$ can be.  We have shown that the cross
section is severely constrained from above by (i) the perturbativity
of the coupling constants up to a certain scale, and (ii) the
consistency of the scale dependence of $\alpha_3$ with that observed
by the LHC.

First, we have pointed out that a model-independent upper bound on
$\sigma( pp\rightarrow S\rightarrow\gamma\gamma)$ can be derived,
taking account of the two requirements mentioned above.  Such a bound is
obtained from the fact that the cross section is related to $\Gamma
(S\rightarrow gg)$ and $\Gamma (S\rightarrow\gamma\gamma)$, and that
the amplitudes for these decay rates are proportional to the $\beta$-function
coefficients of the gauge coupling constants from the fermions and
bosons inside the loop.  We have also calculated such a bound as a
function of the cutoff scale $\Lambda$ and the $m^{\rm eff}_{\rm min}$
parameter which corresponds to the mass scale of the fermions and
bosons inside the loop.  (See Fig.\ \ref{fig:maxGammaAndDiphoton}.)

Then, we have discussed the upper bound on $\sigma (pp\rightarrow
S\rightarrow\gamma\gamma)$ in models with fermion loops, taking into
account the perturbativity of the Yukawa coupling between $S$ and the
new fermions.  For such a study, the particle content should be fixed
to perform the RG analysis.  We have considered seven possible
representations of the fermions with which the fermions can directly
decay into SM particles.  We have introduced $N$ copies of fermions in
the same representation with the universal mass of $m$, and derived
the upper bounds on $\sigma( pp\rightarrow S\rightarrow\gamma\gamma)$.
Among them, the representation of $(\mathbf{3},\mathbf{2},7/6)$ can
give the largest diphoton rate in most of the parameter region.  For
instance, in the case of pseudoscalar, it is shown that $\sigma
(pp\rightarrow S\rightarrow\gamma\gamma)=5$ and $10\ {\rm fb}$ can be
obtained with $m\simeq1000$ and $740\ {\rm GeV}$ ($m\simeq1600$ and
$1100\ {\rm GeV}$) and $N=1$ ($N=2$) when the cutoff scale is
$10^{16}\ {\rm GeV}$ ($10^{10}\ {\rm GeV}$), respectively.  We have
also discussed that such sets of parameters are consistent with the
current constraints on vector-like quarks from the direct search at
the LHC.  In the cases of the other representations, the signal rate
$\sigma (pp\rightarrow S\rightarrow\gamma\gamma)$ is more suppressed,
and a large cutoff scale is impossible at all in some cases.

The present study suggests that, unless the cutoff scale is very low, 
there must exist new particles at TeV scale or lower. 
They should be an important target of the LHC run-2 and other future collider experiments.

\section*{Acknowledgement}
The authors thank Masafumi Kurachi 
for bringing our attention to the running $\alpha_3$ constraint.
This work was supported by Grant-in-Aid for Scientific research Nos.\
23104008 (TM), 25105011 (ME), 26104001 (KH), 26104009 (KJB and KH), 26247038 (KH),
26400239 (TM), 26800123 (KH), and by World Premier International
Research Center Initiative (WPI Initiative), MEXT, Japan.

\end{document}